\documentclass[doublecol]{epl2_AKLF} 

\usepackage{amsmath}
\usepackage{amsfonts}
\usepackage{graphicx}
\usepackage{graphics}

\title{Cold atoms: A field enabled by light}

\author{L. Fallani\inst{1} \and A. Kastberg\inst{2}}
\shortauthor{L. Fallani and A. Kastberg}

\institute{                    
  \inst{1} Department of Physics and Astronomy \& LENS, University of Florence, Via Nello Carrara 1, 50019 Sesto Fiorentino, Italy\\
  \inst{2} Universit\'e Nice Sophia Antipolis, CNRS, Laboratoire de Physique de la Mati\`ere Condens\'ee, UMR 7336, Parc Valrose, 06100 Nice, France
}

\abstract{
Besides being a source of energy, light can also \textit{cool} gases of atoms down to the lowest temperatures ever measured, where atomic motion almost stops. The research field of cold atoms has emerged as a multidisciplinary one, highly relevant, \textit{e.g.}, for precision measurements, quantum gases, simulations of many-body physics, and atom optics. In this focus article, we present the field as seen in 2015, and emphasise the fundamental role in its development that has been played by mastering \emph{light}.  
}

\begin{document}

\maketitle

\section{Introduction}

Cold atom physics has become a mature field of research, but this maturity has been achieved relatively fast. Ideas about mechanical action of light have existed for a long time, and there were preliminary experiments in the late '60s and in the '70s involving cooling and trapping of atoms with light. However, the significant experimental developments gathered momentum in the mid '80s.

There are many conceptual and technological ingredients involved in this research. However, one thing that runs through everything at every stage is \emph{light}. Without the ability to control and detect light with high precision, the field of cold atoms would not exist. This article about cold atoms, inspired by the International Year of Light 2015, is therefore dedicated to \emph{light}.

Traditionally, the field of `cold atoms' is considered a sub-domain of atomic physics. This has historical reasons, but today it could be argued whether it is an ideal designation. In the early days of the field, one driving force was fundamental metrology, and in particular atomic clocks. Moreover, much of the experimental methodology can be traced to `traditional' experimental atomic physics, not least the use of lasers. However, in 2015 the field has become truly multidisciplinary.

Cold atoms play important roles in both fundamental and applied science. Maybe in particular, the field has become an important stage for research in condensed-matter physics, many-body physics and quantum phase transitions. There are also strong connections to molecular physics, plasma physics, nuclear and elementary-particles physics, and even tests of the Standard Model. As a tool, cold atoms are arguably the most important component in the most accurate clocks, but the applications within precision measurements range further, including  sensing and navigation. They are used in space applications, and provide realizations of fundamental quantum information and quantum technologies.

In a way, the field of laser cooling and trapping represented a break with tradition in experimental atomic physics. In the decades following WWII, the experimental evolution was driven towards `big science', with experiments more and more performed at large installations. A cold atom experiment, in contrast, typically takes up modest laboratory space; frequently the experiments are table-top ones. The one thing that made it possible to drive the evolution to more advanced and ambitious explorations, even without bigger machines, was the increased proficiency in handling and detecting \emph{light}.

The particular form of light we use for controlling experiments is laser light. The essential properties of this are its spectral purity, high degree of coherence and the possibility to control the direction and polarization of laser beams. Before cold atoms, it was well established to use laser light to manipulate the \emph{internal} degrees of freedom of atoms. With the advent of \emph{laser cooling} came the first attempts to control the \emph{external} degrees of freedom. As a matter of fact, light carries momentum, which it can transfer to (or dissipate from) atoms. The concept of momentum transfer was the revolutionary ingredient when the field of laser cooling took off. Starting from this simple, yet powerful idea, we learned how to set up combinations of laser fields in order to achieve position-dependent forces for confinement, and velocity-dependent ones for cooling. 

The purpose of this focus article is neither to provide a complete account of the evolution of the field, nor to credit all important contributions. Rather, the objective is to give an overview of the field to a broad audience, from a perspective in 2015. The pioneers of the field have been credited at many occasions, not least with the Nobel prizes in 1997 \cite{Chu1998,Cohen-Tannoudji1998,Phillips1998} and 2001 \cite{Cornell2002,Ketterle2002}, and many references and historical attributions can be found in those Nobel articles. 

Besides the Nobel prizes directly awarded to the field of cold atoms, there are strong connections between the field and the subject matter for other Nobel prizes, in particular those in 2005 and 2012. One of the things that the discoveries, for which those prizes were awarded, have in common with the field of cold atoms is that the most important ingredient in the research is \emph{light}.

\section{A brief history of cold atoms}
\label{field_evolution}

The first successful experiments involving manipulation of external degrees of freedom of atoms came in the decades after the invention of the laser, \textit{e.g}, when atomic beams were deflected by laser light in the early '70s \cite{Picque1972,Schieder1972}. In 1975, there were parallel suggestions \cite{Hansch1975,Wineland1975} published about how to use the velocity dependence of the light-atom interaction in order to narrow the velocity distribution of a sample of either neutral atoms or trapped ions. In these `Doppler cooling' schemes, energy is dissipated by the atoms absorbing radiation, and then re-emitting at a higher photon energy.

\begin{figure}
\centerline{\includegraphics[width=0.85\columnwidth]{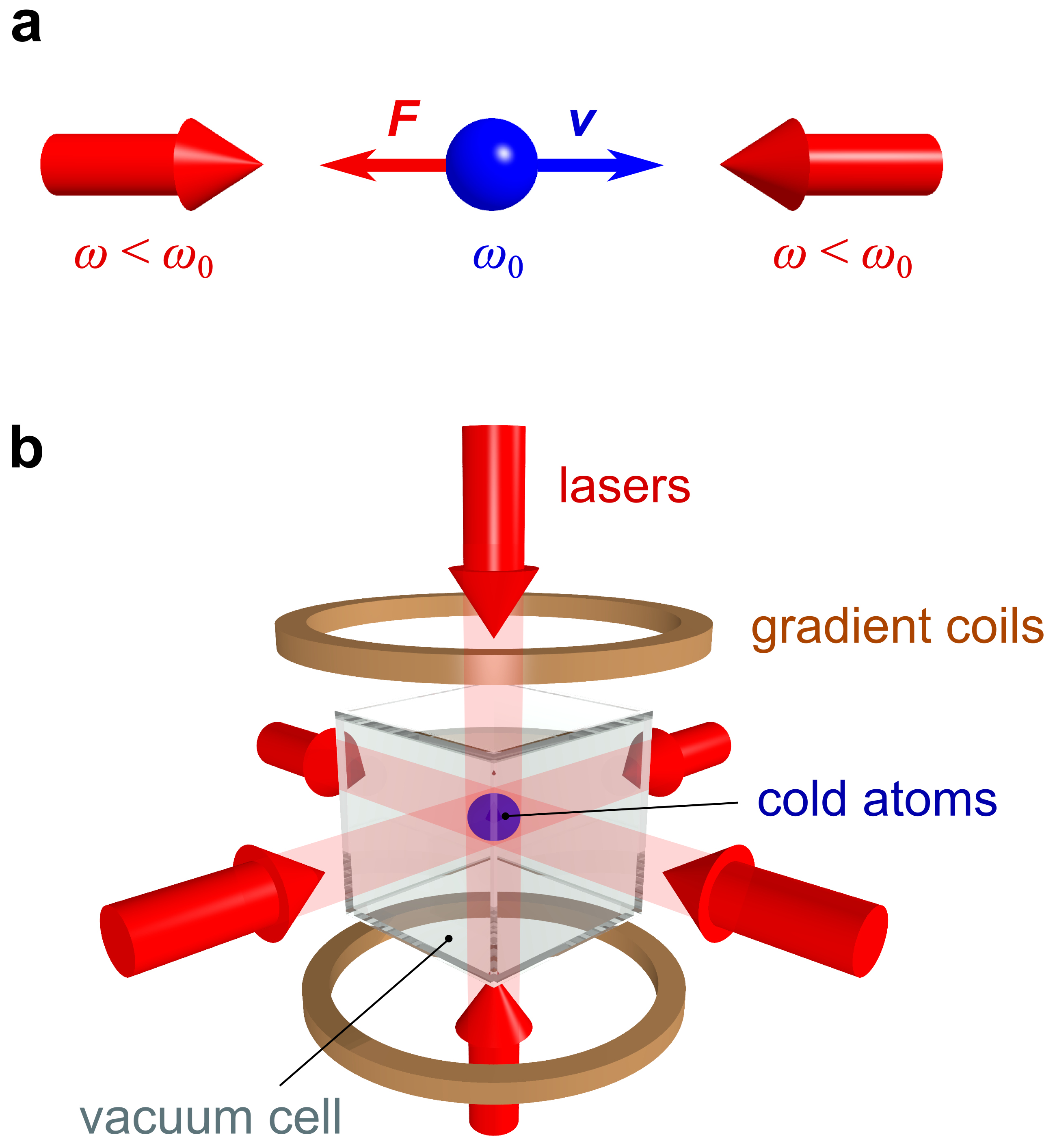}}
\caption{Laser cooling. {\bf a.} An atom is illuminated by opposing laser beams, having a frequency $\omega$ slightly lower than the transition frequency $\omega_0$. Due to the Doppler effect, a moving atom preferentially absorbs photons from the counterpropagating beam, and the resulting radiation pressure force is opposed to the atom velocity. {\bf b.} In a magneto-optical trap the combination of a magnetic field gradient and counterpropagating lasers produces a restoring force, in addition to the cooling force, and provides trapping of the atoms.}
\label{fig1}
\end{figure}

The basic idea of Doppler cooling is to use counterpropagating laser beams tuned slightly below an atomic resonance, as shown in Fig. \ref{fig1}a. An atom feels a stronger resonant radiation pressure force from the laser propagating contrary to its motion, and thereby a velocity-dependent force is obtained. Early experiments exploiting these ideas were made with ions (see \textit{e.g.} \cite{Wineland1978} and \cite{Wineland1979}). 

\subsection{Bringing neutral atoms to a halt}
In the early '80s, lasers were being used to manipulate atom beams, with the goal of bringing the atoms to a halt. To stop a thermal beam, the intuitive method is to apply resonant radiation pressure with an opposing laser. This force can be substantial; in the case of Na atoms one gets a maximum acceleration of $10^5 g$. Thus, it should be possible to stop atoms emerging from an oven in a meter or so. However, since the laser beam must compensate for the Doppler shift of the fast atoms, the decelerated atoms are shifted out of resonance. One way to solve this problem \cite{Prodan1985} is to use a varying longitudinal magnetic field applied along the atomic beam. This gives a position-dependent Zeeman shift, and when well designed, atoms can be stopped. 

With stopped atoms, it proved possible to apply a three-dimensional Doppler cooling scheme. That is, a configuration of three pairs of counterpropagating, red detuned, laser beams; one along each coordinate axis. This configuration was named `optical molasses' \cite{Chu1985}, due to the viscous nature of the effect of the light. Albeit the atoms are not confined, the diffusion time through the molasses is long enough to enable diagnostics. 

The limitation to Doppler cooling comes from the random nature of spontaneous emission and therefore the lowest possible temperature is inversely proportional to the excited state lifetime. For typical atomic systems, this low-temperature limit is some hundreds of microkelvin.



\subsubsection{Sub-Doppler cooling}
A few years after the demonstration of optical molasses, there was a breakthrough for more advanced cooling methods. This started with a refinement in how kinetic temperatures were measured. With a new type of thermometry \cite{Lett1988}, an atomic cloud was released by turning off the laser. With the atoms in free fall, the cold cloud expands and eventually falls through a thin sheet of light. From the fluorescence, the expansion can be derived, and the velocity distribution at time of release can be calculated. The extraordinary result in these measurements was that the measured temperatures were significantly lower than the predicted theoretical limit. 

The efficient cooling soon got its explanation \cite{Dalibard1989,Ungar1989} and the key is state degeneracy. In two-level Doppler cooling, the radiative lifetime is the only involved timescale. In a degenerate multi-level system the population distribution may evolve on a slower timescale, causing a lag between external and internal evolutions. If the compound light field has proper polarisations, the steady-state temperature can get significantly lower than the `Doppler limit', and get as low as a few microkelvins. The limit for these `polarisation-gradient cooling' schemes is set by the momentum related to the recoil of a single emitted photon.

\begin{figure}
\centerline{\includegraphics[width=0.9\columnwidth]{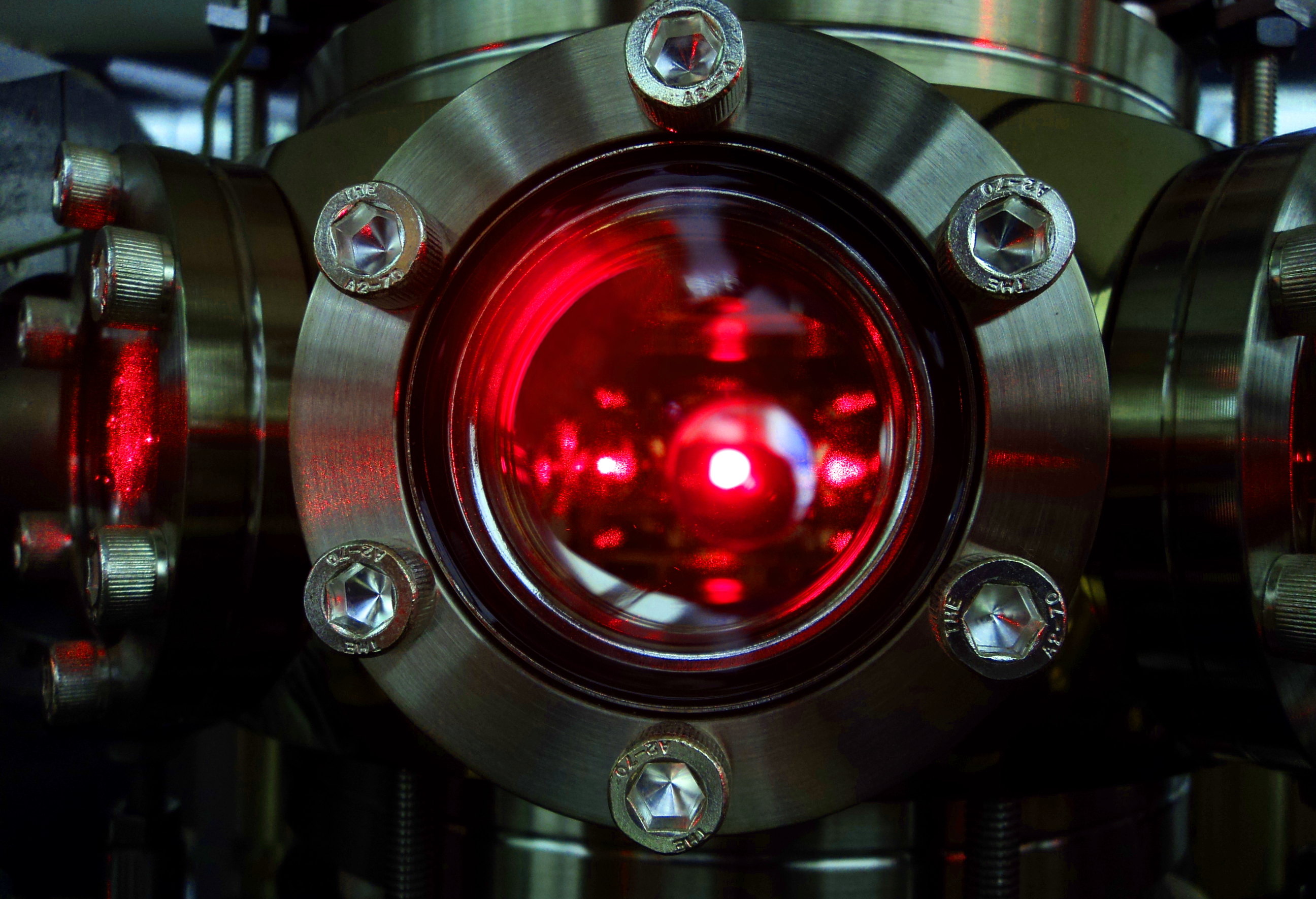}}
\caption{A laser cooling experiment. Inside the vacuum chamber a ball of $\sim 10^9$ cold $^6$Li atoms (the white spot in the centre), at a temperature of $\sim 1$ mK, glows while it scatters photons from the beams of a magneto-optical trap (courtesy of G. Roati, LENS).} 
\label{fig2}
\end{figure}

\subsection{Trapping 
atoms}
In many early laser cooling schemes, atoms were not confined. An efficient way to trap them was lacking. Nowadays, neutral atom trapping is well established, and various techniques are used.

One successful approach is based on off-resonance light. When this interacts with an atom, the energy levels are perturbed via the light shift. The sign of this shift depends on the sign of the detuning, and its size is proportional to the light intensity. This means that a local energy minimum can be designed in an inhomogeneous light field. Provided that the detuning is large, the resulting trap is essentially conservative. Such `dipole traps' have been used to trap also larger objects than atoms, like living bacteria, in `optical tweezers' \cite{Ashkin1986,Moffitt2008}. Using several beams, a periodic pattern of atom traps can also be created; a so called `optical lattice'.

Static magnetic fields have also been successfully used to make conservative atomic traps. For example in a local field minimum, a low-field seeking state may be trapped.


\subsubsection{The MOT}

The most widely used source of cold atoms is the MOT, \textit{i.e.} the Magneto-Optical Trap, first demonstrated in 1987 \cite{Raab1987}. In this trap, an inhomogeneous magnetic field is combined with three pairs of counterpropagating laser beams, yielding a restoring light pressure force, as shown in Fig. \ref{fig1}b.

In a MOT, atoms with velocities as high as several tens of m/s can be captured. This means that the low velocity tail of a room temperature sample can be trapped, which in many cases eliminates the need for an initial beam apparatus. Thereby the experimental set-up can be significantly simplified, and even miniaturised. A photo of a typical experimental realization is showed in Fig. \ref{fig2}.

\subsection{Atom optics}
With atoms moving as slowly as some mm/s, their wave nature becomes important, and the small momentum dispersion makes the sample quasi-monochromatic. This opens up possibilities for {\it matter wave optics}. Diffraction with cold atoms \cite{Gould1986} and reflection from an `evanescent wave mirror' \cite{Balykin1988} were demonstrated in the '80s. Subsequently, various atom interferometers have been demonstrated \cite{Martin1988,Borde1989,Rasel1995}. The most accurate current realisations of the second are in fact atom interferometers (as described in a later section).

\subsection{Bose-Einstein condensation}
An ambitious idea in the early days of cold atom physics was to create a Bose-Einstein condensate (BEC). This would mean reaching a phase-space density of the order of one, \textit{i.e}, a de~Broglie wavelength as large as the typical separation between atoms. This was achieved in 1995 \cite{Anderson1995,Bradley1995,Davis1995}.

\begin{figure}[b!]
\centerline{\includegraphics[width=0.85\columnwidth]{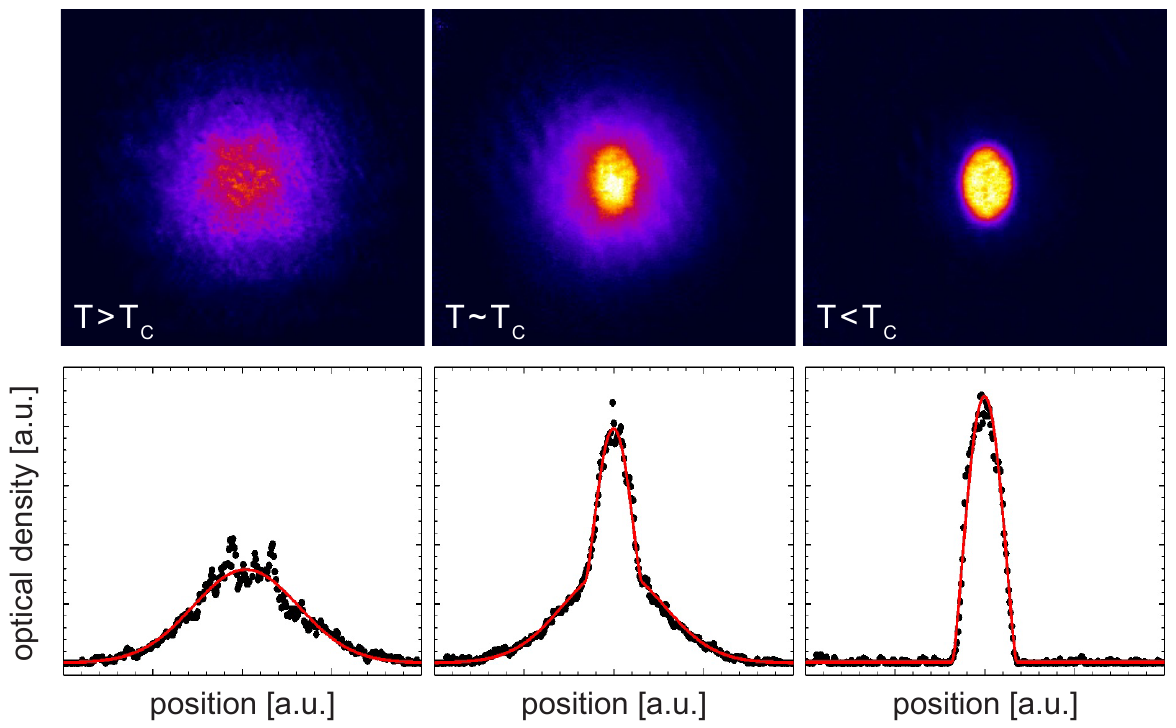}}
\caption{Bose-Einstein condensation. Velocity distribution of an ultracold cloud of $^{87}$Rb (recorded with time-of-flight absorption imaging) for temperatures $T$ above, around, and below the BEC critical temperature $T_\mathrm{c}\sim 100$ nK. The BEC transition is evidenced by a change in the aspect ratio of the cloud (see images above), and by a change of the cloud shape from a Gaussian to an inverted parabola (see integrated densities below) (data from LENS \cite{Fallani2005}).}
\label{fig3}
\end{figure}

To realise a BEC, it was necessary to beat the recoil limit, \textit{i.e.}, to use a cooling method that does not carry away energy by random photon recoils. The solution was evaporative cooling, and still all BEC experiments involve versions of this method. It is conceptually simple; atoms are pre-cooled, then loaded into a conservative trap, and as a final cooling stage the trap potential is gradually lowered. This allows highly energetic atoms to `boil off', and if the density is high enough the remaining atoms thermalise to a lower temperature. Figure \ref{fig3} shows velocity distributions before and after BEC has been achieved.

Two properties of BECs that make them unique platforms for a wide range of fundamental studies and applications are coherence and superfluidity. Among things that have been made possible are experiments with truly monochromatic matter waves. Even an analogue to a laser, an `atom laser', has been realised \cite{Mewes1997,Andrews1997}. Matter wave optics is also intrinsically suitable for non-linear optics, because of the atom-atom interaction. The connections to quantum fluids and superfluidity have been demonstrated, for example in the observation of quantum vortices \cite{Madison2000,Abo-Shaeer2001}.


\section{Current status of cold atom physics}
The field of cold atoms has evolved rapidly, and this is still going on. In the following, we will take a closer look at some of the most intense research directions that have been opened by cold atoms.

\subsection{Quantum matter}
The achievement of BEC opened up a wealth of possibilities for fundamental and applied science. A completely new field of research has emerged, in which ultracold atoms are used as elementary components in the investigation of fundamental properties of quantum matter \cite{inguscio2013}. Significantly, it has become possible to make unprecedented observations of the effects of quantum statistics, using both bosonic and fermionic samples.

After the achievement of quantum degeneracy in bosonic gases, several research groups tried to reach the same regime for fermions. Initial efforts were hampered by the lack of thermalisation in evaporative cooling, caused by the absence of ultracold collisions between identical fermions (imposed by the symmetrisation principle). Eventually, a spin mixture of $^{40}$K potassium atoms was successfully cooled far below the Fermi temperature, where a quantum degenerate Fermi gas forms \cite{demarco1999}. Subsequently, it became possible to cool different atomic species simultaneously, allowing the realization of various ultracold quantum mixtures (Bose-Fermi, Bose-Bose and Fermi-Fermi).

After early studies of the properties of weakly-interacting Bose-Einstein condensates and Fermi gases, large efforts were directed on how to make the ultracold gases strongly interacting \cite{bloch2008}. Two fundamental tools have been developed to answer this question.

\subsubsection{Feshbach resonances}

One is the use of magnetic fields to control atom-atom interactions. By tuning the value of a static, uniform magnetic field across a so-called Feshbach resonance it is possible to control the coupling between internal states in atom-atom collisions \cite{chin2010}. As a consequence, the effects of the collision are dramatically changed. The interaction strength can be cancelled, maximised, or it can even change sign -- from repulsive to attractive and vice versa. 

This has allowed fundamental studies of strongly-interacting quantum matter in regimes where many-body quantum correlations are dominant. It has enabled the study of `fermionic superfluidity', in which two fermions in different spin states can form pairs and condense to a superfluid state \cite{inguscio2008}. In particular, it has allowed a detailed study of fermionic superfluidity across the whole BEC-BCS crossover, which continuously connects the phenomenon of Bose-Einstein condensation (when fermions form tightly-bound molecules with bosonic character) to the mechanism of BCS pairing (when they form weakly-bound Cooper pairs), which is responsible for the lack of dissipation in ordinary superconductors.

\begin{figure}[b!]
\centerline{\includegraphics[width=0.85\columnwidth]{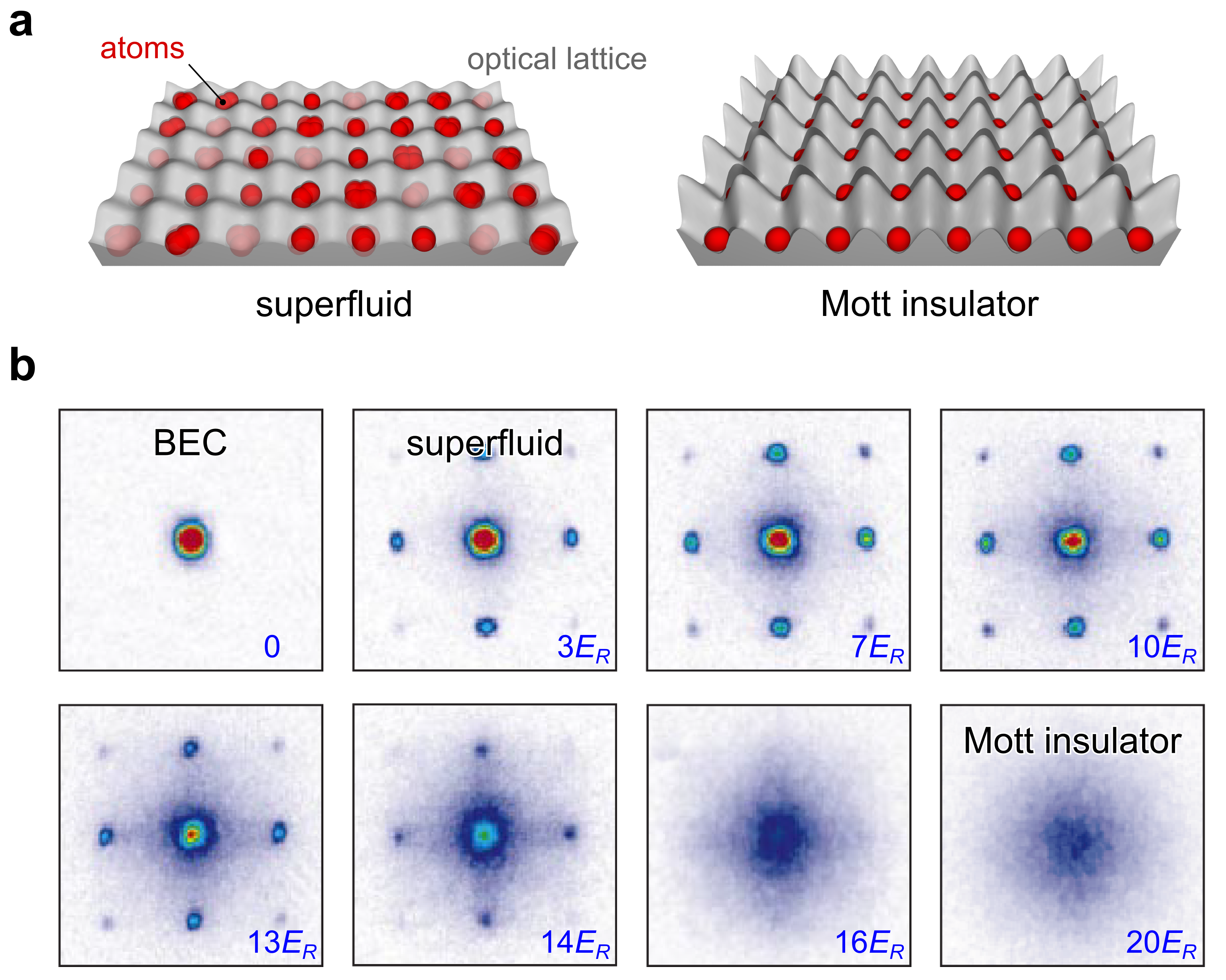}}
\caption{Quantum simulation in optical lattices. {\bf a.} The ground state of a system of bosons in an optical lattice changes from a superfluid at weak interactions to a Mott insulator at large interactions, with localization of single atoms in the lattice sites. {\bf b.} The quantum phase transition is detected by measuring the phase coherence of the atomic state, probed with matter-wave interference in a time-of-flight experiment (picture from Ref. \cite{greiner2002}).}
\label{fig4}
\end{figure}

\subsubsection{Optical Lattices}
The second tool was provided by light in form of optical lattices. These are standing waves of light produced by the interference of two or more intersecting laser beams. The resulting intensity modulation produces an array of microscopic traps, located at the nodes (or antinodes) of the laser field, in which atoms can be tightly trapped. The use of optical lattices has opened a new exciting direction of research, stimulated by the strong analogy with solid-state physics. 

Atoms moving in this artificial `crystal of light' obey the same laws as electrons moving in an ideal crystalline solid, since both can be described as quantum particles interacting with a periodic potential. The extremely long coherence times, and the absence of defects in the optical lattice, provide the ideal environment for investigating the transport of quantum particles in periodic structures, as studied in the late '90s with cold atoms \cite{raizen1997} and later with quantum degenerate gases \cite{morsch2006}.

\subsubsection{Quantum simulations with optical lattices}
In a quantum simulation perspective, optical lattices have been used to experimentally realize paradigmatic condensed-matter lattice models in which the Hamiltonian parameters can be precisely controlled and tuned \cite{lewenstein2012}. In an experiment with a three-dimensional optical lattice \cite{greiner2002} the Bose-Hubbard model, which describes repulsively interacting bosons in a lattice, was realized in a clean way. This provided the first evidence of a quantum phase transition from a bosonic superfluid to a Mott insulating state, as the atom-atom interaction energy was made larger than the kinetic `hopping' energy, as illustrated in Fig. \ref{fig4}.

Later work with fermions allowed the study of different quantum phases, ranging from metallic states to band insulators and fermionic Mott insulators \cite{jordens2008,schneider2008}. The realization of controlled disorder, by using speckles or quasi-periodic light patterns, also allowed the observation of an insulating transition driven by the phenomenon of Anderson localization \cite{billy2008,roati2008}. 
Recently, the development of high-performance imaging systems has allowed the detection of single atoms trapped in optical lattices \cite{bakr2009,sherson2010}, allowing new exciting perspectives.

\subsubsection{Low-dimensional systems}
Trapping ultracold atoms in optical lattices also allows the production of low-dimensional quantum gases. Two-dimensional (2D) systems can be created by using a deep one-dimensional (1D) optical lattice which freezes the atomic motion along one direction. This has allowed the observation of fundamental phenomena such as the Berezinski-Kosterlitz-Thouless transition experienced by an interacting 2D Bose gas \cite{hadzibabic2006}. Deep 2D optical lattices allow the realization of 1D quantum gases. Such atomic quantum wires are interesting because quantum correlations are most pronounced in one dimension, resulting in phenomena such as the `fermionisation' of bosonic 1D gases \cite{kinoshita2004,paredes2004}.

\subsubsection{Long-range and 'synthetic' interactions}
Great experimental effort is currently directed towards ultracold `atomic' systems that exhibit long-range interactions, and which could grant access to a vast scenario of quantum phenomena. These systems include quantum gases of highly-magnetic transition metals or lanthanide elements, recently brought down to quantum degeneracy \cite{griesmaier2005,lu2011,aikawa2012}, in which the magnetic dipole-dipole interaction (usually neglected in alkaline or alkaline-earth atoms) has a dominant effect. Other interesting systems are polar molecules formed by atoms of different species. These heteronuclear molecules have a permanent electric dipole moment, which is also responsible for long-range dipole-dipole interactions. 

On the one hand, laser cooling of simple molecules is currently under investigation, and despite difficulties caused by the complex structure of molecular spectra, initial results have been encouraging \cite{Zhelyazkova2014}. Moreover, experiments are being performed with heteronuclear diatomic molecules, which can be formed from ultracold atoms, either by using Feshbach resonances or by using light in a photoassociation process \cite{Jones206}. 

Further perspectives are opened by the possibility of using light to engineer new kinds of interactions for neutral atoms. For instance, under appropriate experimental schemes, light-matter interaction can imprint a phase on the atomic wave function which is formally equivalent to the Aharonov-Bohm phase experienced by a charged particle moving in a magnetic field \cite{dalibard2011}. These schemes have allowed the production of synthetic magnetic fields for effectively charged particles \cite{lin2009}, with the perspective of investigating quantum Hall physics, and the realization of spin-orbit coupling in ultracold quantum gases \cite{lin2011}.

\subsection{Precision measurements}
The interaction of light with atoms has always provided precise measurements that have in turn stimulated the development of fundamental theories, and resulted in stringent tests of their validity. Precision atomic spectroscopy on simple and `calculable' atomic systems such as hydrogen \cite{parthey2011} now allows extreme tests of quantum electrodynamics and the precise determination of fundamental constants. Recent measurements of the Lamb shift in muonic hydrogen allowed a precise determination of the proton charge radius \cite{pohl2010}, opening a puzzle - still unsolved - caused by the inconsistency of the measured value with previous experimental determinations.

\subsubsection{Atomic clocks}
\begin{figure}[t!]
\centerline{\includegraphics[width=0.9\columnwidth]{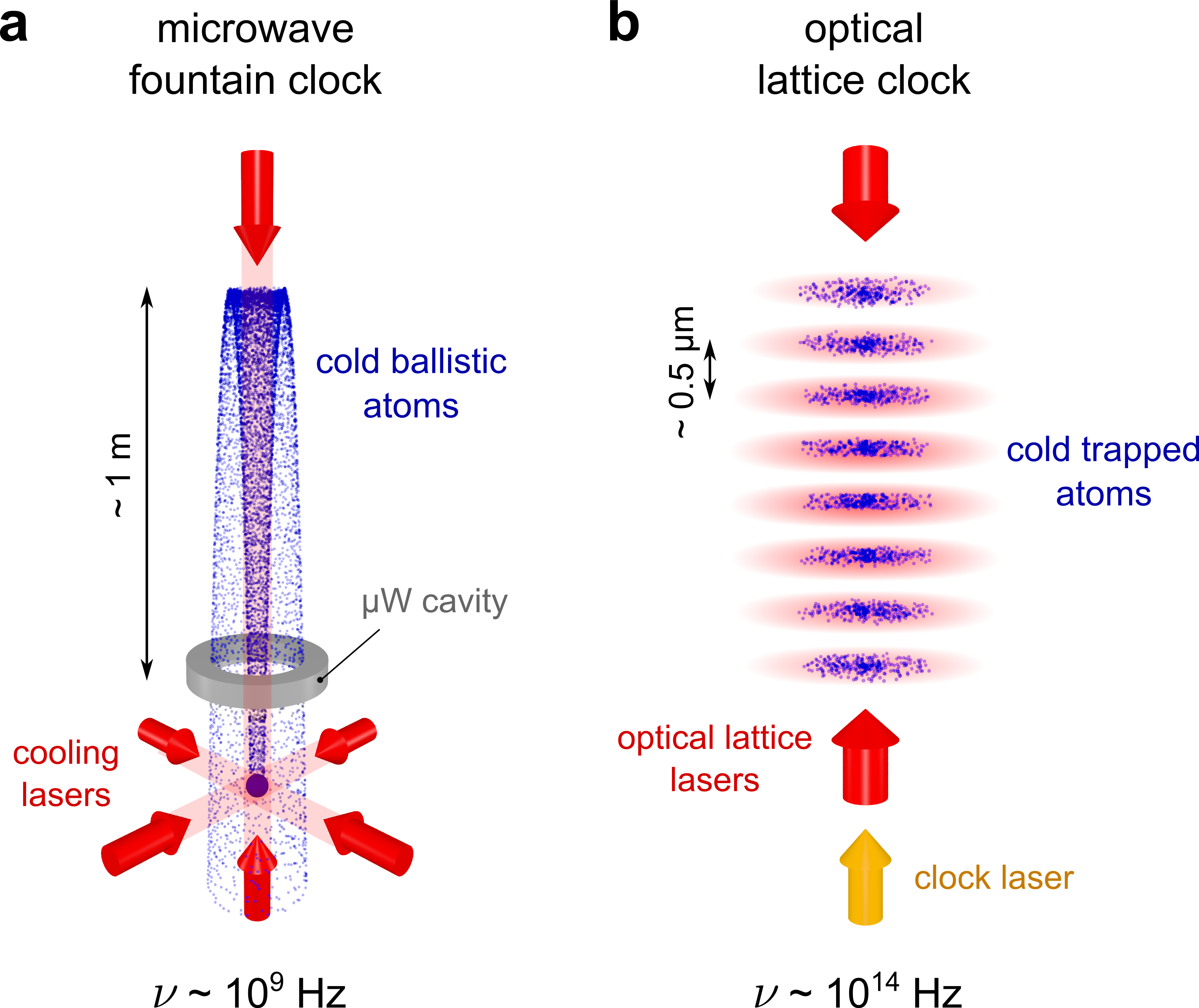}}
\caption{Laser-cooled atomic clocks. {\bf a.} In state-of-the-art microwave clocks an atomic fountain provides long interrogation times on the order of one second. {\bf b.} In the new generation of optical clocks the atoms are trapped in an optical lattice, which freezes the atomic motion and provides Doppler- and recoil-free measurements.}
\label{fig5}
\end{figure}

The definition of the second is based on the frequency of a microwave transition connecting two hyperfine levels of caesium ($\sim$9.2 GHz). One of the main limits to the precision of this spectroscopic measurement is given by the finite microwave-atom interaction time. The invention of laser cooling has enabled new atomic clocks, in which it is possible to achieve much longer interaction times than those achievable in traditional set-ups. Nowadays, the most accurate caesium clocks are built around an `atomic fountain' concept, in which a slow and collimated sample of atoms is first launched upwards about a meter, whereafter they fall down (as sketched in Fig. \ref{fig5}a). In this way, interrogation times of the order of one second are possible\cite{wynands2005}.

In the last decade a new generation of atomic clocks, based on optical transitions rather than on microwave ones, has been demonstrated \cite{poli2013,ludlow2014}. These optical atomic clocks operate on ultranarrow transitions in laser-cooled ions or ultracold two-electron atoms, providing a faster and more accurate frequency reference ($\sim$500 THz). The atoms are tightly trapped in the so-called Lamb-Dicke regime, either in electrodynamic traps (for ions) or optical lattices (for neutral atoms). In this limit the interaction with light does not change the motional state of the trapped particles and the transition frequency is not affected by any detrimental Doppler or recoil shift (a neutral-atom optical lattice clock is schematically sketched in Fig. \ref{fig5}b).

Developments like the invention of the optical frequency comb for the precise measurement and transfer of optical frequencies \cite{diddams2000,cundiff2003}, and sub-Hz linewidth lasers  \cite{martin2011}, have allowed a spectacular progress in the performance of these clocks. Uncertainties as small as $10^{-18}$ have been reported \cite{hinkley2013,bloom2014,ushijima2015} -- an improvement of about two orders of magnitude over the precision of the microwave atomic clocks on which the current definition of the second is based.

\subsubsection{Sensing and fundamental physics}
Precise measurements with cold atoms are not limited to clocks. Cold atoms can be used as de Broglie matter waves in atom interferometers, in which the interference of atomic wave functions is used to detect phase shifts, as in conventional interferometers with light \cite{cronin2009}. In the most successful interferometric schemes, the role of light and matter is basically inverted and light constitutes the atom-optical devices (such as mirrors, beam-splitters and gratings) that are used to manipulate and control matter waves. Atom interferometers have allowed the measurement of fundamental constants such as the fine-structure constant \cite{Webb2011}, and proved to be precise sensors for the local gravitational acceleration \cite{peters1999}, the Newton gravitational constant \cite{fixler2007,rosi2014}), and also for inertial forces \cite{gustavson1997}.

\subsection{Quantum information}
Cold atoms are also a valuable resource in the field of quantum information. Generally, this follows one of two main branches. On the one hand, ensembles of cold atoms are used to reproduce the behaviour of systems or models which cannot be handled by classical computers: this subfield, quantum simulation, was covered in a preceding section. On the other hand, cold atoms and trapped ions are used as a platform for processing elementary bits of quantum information, with the long-term goal of building a quantum computer.

There are many different physical systems that are explored in connection with quantum computing. The advantages with cold atomic systems are the excellent isolation against ambient perturbations, and the level of quantum control. In terms of realising a scalable quantum computer, cold trapped ions are arguably currently more promising than neutral atoms \cite{Monroe2014}. Chains of ions have been entangled, and there have been several demonstrations of quantum gates (\textit{cf.} \cite{Roos2014}). There are also a number of suggestions for using cold neutral atoms in optical lattices as a quantum register \cite{Mandel2003}, and some two-qubit gate operations have been demonstrated (\textit{e.g.} \cite{Reiserer2014}).

\section{Outlook}
Thirty years after the first three-dimensional laser cooling, cold atoms have become an established and multidisciplinary field of research. Many different research routes have been opened and we expect much more developments in the future.

In the case of interacting many-body quantum systems, we foresee many breakthroughs in the future due to increasingly more advanced techniques of quantum control and quantum engineering. Developments in the realization of synthetic quantum systems with single-atom detection/manipulation control and the implementation of long-range interactions, bring the dream of realizing a universal quantum computer closer to reality. Experiments with cold gases in optical lattices are already providing dedicated quantum computers, in the form of tabletop quantum simulators, which already outperform the capabilities of algorithms run on classical supercomputers. 

Quantum simulation experiments are providing solutions to problems that are classically unsolvable, and they are likely to offer theoretical insight into key open problems of physics, ranging from the nature of the fractional quantum Hall effect to high-$T_\mathrm{c}$ superconductivity. Cold atoms physics will allow us to study quantum matter under extreme conditions, without other analogues in laboratories. This could even be extended to complex phenomena in high-energy physics, including the properties of exotic fermionic superfluids, relevant for the behaviour of nuclear matter inside neutron stars, and the physics of matter coupled to dynamical gauge fields, with the long-term goal of realizing a cold-atom simulator of distinctive aspects of quantum chromodynamics.

In the case of fundamental metrology, the precision of optical atomic clocks is improving at the remarkable rate of roughly one order of magnitude every 4 years. These clocks are still operating far from fundamental limits imposed by quantum mechanics, so we expect significant further progress in the future. The increase in performance of optical clocks is connected with the possible future redefinition of the SI unit of time on the basis of an optical standard.

With this perspective in mind, metrological optical fibre links are being realized in order to allow remote comparisons between different clocks, also on a continental scale. The development of better and better interconnected clocks, in combination with the operation of precise atom interferometers, will provide us with new sensitive tools for studies of fundamental physics, for example, for the detection of gravitational waves. These applications could be enhanced by the operation of cold atoms devices in space. Furthermore, high-precision spectroscopy of trapped samples of anti-matter will offer us stringent tests of the validity of the equivalence principle.

The research on cold atoms is providing answers to open questions concerning the properties of the constituents of matter. Even more, it is addressing deeply fundamental issues about our understanding of the natural world. Not least, it is enabling a new form of quantum technology which uses quantum mechanics as a tool for measurements, sensing and computing applications. Of all the developments that have made this possible, the most important one has been our increased understanding about the properties and the handling of light, and its interaction with matter. Thus, we can safely say that the science of cold atoms has truly been \emph{enabled by light}.

%
%

\end{document}